\documentclass[10pt, conference]{IEEEtran}
\IEEEoverridecommandlockouts
\usepackage{algorithm,epsfig,cite,stfloats,bm,bbm,amssymb,amsmath,color, subfigure,graphics,extpfeil,amsthm, caption, microtype}
\usepackage[T1]{fontenc}
\usepackage{multirow}

\usepackage[noend]{algorithmic}
\def\BibTeX{{\rm B\kern-.05em{\sc i\kern-.025em b}\kern-.08em
    T\kern-.1667em\lower.7ex\hbox{E}\kern-.125emX}}
\begin{document}

\title{Digital Twin-Assisted Robust and Adaptive Resource Slicing in LEO Satellite Networks}
\author{\IEEEauthorblockN{Mingcheng He$^*$, Huaqing Wu$^\dag$, Conghao Zhou$^*$, Shisheng Hu$^*$, Zhixuan Tang$^*$, and Weihua Zhuang$^*$}
\IEEEauthorblockA{$^*$Department of Electrical and Computer Engineering, University of Waterloo, Canada}
\IEEEauthorblockA{$^\dag$Department of Electrical and Software Engineering, University of Calgary, Canada}
\IEEEauthorblockA{E-mails: \{m64he, c89zhou, s97hu, z32tang, wzhuang\}@uwaterloo.ca, huaqing.wu1@ucalgary.com}
}

\maketitle
\begin{abstract}
Resource slicing in low Earth orbit satellite networks (LSN) is essential to support diversified services.
In this paper, we investigate a resource slicing problem in LSN to reserve resources in satellites to achieve efficient resource provisioning.
To address the challenges of non-stationary service demands, inaccurate prediction, and satellite mobility, we propose an adaptive digital twin (DT)-assisted resource slicing scheme for robust and adaptive resource management in LSN.
Specifically, a slice DT, being able to capture the service demand prediction uncertainty through collected service demand data, is constructed to enhance the robustness of resource slicing decisions for dynamic service demands.
In addition, the constructed DT can emulate resource slicing decisions for evaluating their performance, enabling adaptive slicing decision updates to efficiently reserve resources in LSN.
Simulation results demonstrate that the proposed scheme outperforms benchmark methods, achieving low service demand violations with efficient resource consumption.
\end{abstract}
\setlength{\textfloatsep}{2mm}

\section{Introduction}

The 6G networks are envisioned to provide reliable, low-latency, and ubiquitous communication to support diversified network services \cite{shen2021holistic}.
Low Earth orbit (LEO) satellite networks (LSN) are considered as indispensable components of 6G networks, with their extensive communication coverage and dense deployment, to enlarge the limited communication coverage of terrestrial networks and alleviate network congestions, thereby supporting multifarious and ubiquitous applications \cite{azari2022evolution, wu2024network}.

\begin{figure}
    \centering
    \includegraphics[width=0.38\textwidth]{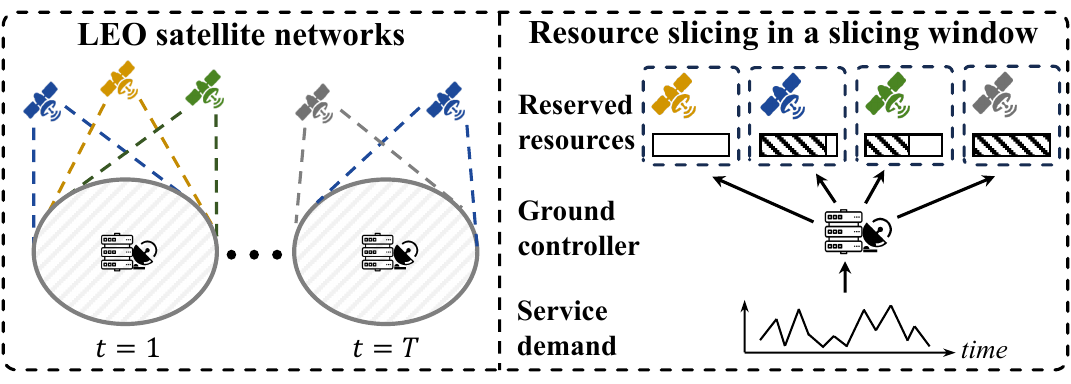}
    \caption{Resource slicing architecture in LSN.}
    \label{fig:architecture}
\end{figure}

As an innovation in service-oriented network management techniques for the 5G era, resource slicing will continue to play a pivotal role in satisfying service requirements of diversified services via efficient resource provisioning \cite{wijethilaka2021survey}. 
Different from resource slicing in 5G targeting fixed network infrastructures such as base stations, resource slicing in LSN needs the construction of virtual networks, referred to as slices, upon different satellites with high mobility while satisfying diverse service-level agreements for different services in any target area.
As shown in Fig. \ref{fig:architecture}, in LSN, the ground controller in an area is responsible for managing network slices in a large time scale, named as a slicing window, by reserving resources in satellites serving the area to accommodate service demands.
Considering different service durations during which satellites cover a given area, the resource slicing decision is first made at the beginning of each slicing window, and then executed by reserving resources for the slice in each satellite when the satellite first covers the area within the slicing window.

There exist some research efforts on resource slicing in satellite networks, by considering the heterogeneity of satellite and terrestrial networks to satisfy different QoS requirements in \cite{de2020qos} and developing a scalable resource slicing scheme to coordinate multiple cells for resource efficiency in \cite{he2024resource}.
Despite these research efforts, realizing resource slicing in LSN still meets the following two challenges.
First, due to the mobility of satellites, the number of satellites covering a designated area varies over time, resulting in inconsistent available resources for network services.
Meanwhile, time-varying positions of satellites result in differentiated communication capabilities due to different channel conditions.
Second, the service demand of LSN is time-varying, and such temporal variations may be non-stationary, thereby challenging the accurate prediction of service demand for proactive resource slicing.
Unlike resource slicing in terrestrial networks, where the reserved resources are generally fixed during the slicing window, resource slicing in LSN has more flexibility.
This flexibility stems from the ability of LSN to dynamically adjust the reserved resources from satellites to accommodate dynamic service demands due to the time-varying available resources from LSN.
Despite the flexibility, efficient resource slicing is still challenging, since the reserved resources from satellites can result in an accumulated impact of under-/over-provisioning of resources considering the inaccurate service demand prediction in the slicing window.  \looseness=-1

Slice DT, as a digital representation of a network slice for the service, attracts wide attention, which can facilitate resource slicing with a data-driven approach \cite{shen2021holistic, khan2022digital}.
To deal with the aforementioned challenges, in this paper, we propose an adaptive digital twin (DT)-assisted resource slicing (ADTRS) scheme for resource provisioning in LSN.
Our objective is to efficiently reserve resources in LSN for the target area to minimize the overall costs related to resource usage and delay, taking into account the dynamic network environment and the uncertainty in service demand prediction.
To achieve the objective, a chance-constrained programming problem is formulated to find robust and adaptive resource slicing decisions.
Then, a slice DT is constructed in the ground controller to depict the dynamic service demand and slicing performance.
Specifically, in the slice DT, we develop a feature extraction module, a prediction module, and a distribution fitting module to handle the uncertainty of service demand prediction.
These modules operate by periodically collecting service demand information of the target area from the ground controller.
By processing the collected data, the designed modules in the slice DT can predict service demand with information on prediction uncertainty to allow robust resource slicing decision-making.
Moreover, an emulation module is devised in the slice DT to evaluate the performance of resource slicing decisions in dynamic environments, facilitating adaptive decision updating in LSN.
Extensive simulations demonstrate that the proposed scheme can outperform benchmark algorithms in terms of low service demand violation and high resource efficiency. \looseness=-1

\section{System Model and Problem Formulation}
\subsection{Network Scenario}
As shown in Fig. \ref{fig:architecture}, consider an LSN network where multiple LEO satellites can jointly serve a target area, with the target area being equal to the coverage of a single satellite beam.
Denote the set of satellites in a constellation by $\mathcal{S}$, each with available bandwidth resources denoted by $B$.
Specifically, LEO satellites are equipped with steerable antennas that can always point to the target area when the elevation angle constraint is met.
A ground controller is deployed in the target area to manage resources from different satellites serving the area.

The resource slicing in LSN operates in a time-slotted system. 
The time is separated into a number of slicing windows $w \in \mathcal{W} = \{1,2,...,W\}$ with each slicing window consisting of $T$ time slots, where each time slot $t \in \mathcal{T} = \{1,2,...,T\}$ has the same duration $\tau$.
For simplicity, let $(w, t)$ represent the $t$-th time slot in slicing window $w$. 
We consider that the position of each satellite is fixed during each time slot and varies over time slots.
Let $a^s_{w, t} = 1$ denote satellite $s$ that can serve the area at time $(w, t)$; otherwise, $a^s_{w, t} = 0$.
The set of satellites involved in serving the area in slicing window $w$ can be represented by $\mathcal{S}_{w} = \{s \mid \sum_{t \in \mathcal{T}} a^s_{w, t} > 0, \forall s \in \mathcal{S}\}$.

In this paper, we consider a network slice in LSN for the target area to serve downlink delay-tolerant services.
Let $l_{w, t}$ denote the actual service demand of the slice in the target area (in packets per second) at time $(w, t)$ following the probability distribution $\mathcal{F}_{w, t}(\mathbf{x}_{w, t})$, where $\mathbf{x}_{w, t}$ is the parameter set of the distribution.
We assume that the service demand distribution is fixed within each time slot. \looseness=-1

\subsection{Communication Model}

To establish a network slice of the service in the target area, bandwidth resources need to be efficiently reserved from different satellites to accommodate service demands from the target area.
Let $b^{s}_{w}$ denote the proportion of reserved resources out of all available resources in satellite $s$ for slicing window $w$. \looseness=-1

For a long-term resource reservation problem, we consider the channel gain to be mainly affected by the average communication distance between users and satellites.
Given the averaged distance between the central point of the area and satellite $s$ at time $(w, t)$ by $d^s_{w, t}$, the channel gain between users and satellite $s$ can be expressed as $h^s_{w,t} = (d^s_{w, t})^{-\delta}$, with the pathloss exponent $\delta$.
In this case, the transmission data rate measured in packets per second can be given by
\begin{equation}
    R^{s}_{w, t} = \frac{a^s_{w, t} b^{s}_{w} B}{\kappa} \log_2 \left(1 + \frac{P h^s_{w,t}}{\sigma^2} \right),
\end{equation}
where $\kappa$ is the packet size and $P$ is the transmission power of each satellite, 

To model the service of each satellite to users in the target area, we adopt a First-Input-First-Output (FIFO) queue to illustrate the transmission process in each satellite.
However, due to stochastic service demands, it is difficult to satisfy the deterministic QoS requirement for the network slice. 
In this case, effective capacity emerges as a valuable approach \cite{wu2003effective} to model the statistical delay performance, which quantifies the maximum service demand that can be accommodated by reserved resources.
Effective capacity facilitates the evaluation of delay-bounded violation probability that delay $D$ exceeds the queue's delay bound $D^{s, \mathrm{queue}}_{w, t}$, presented by
\begin{equation}
    P(D \geq D^{s, \mathrm{queue}}_{w, t}) \approx e^{-\theta C(R^{s}_{w, t}, \theta) D^{s, \mathrm{queue}}_{w, t}},
    \label{eq:delay_violation}
\end{equation}
where $C(R^{s}_{w, t}, \theta)$ is the effective capacity, and $\theta$ is the QoS exponent representing the exponential decay rate of the QoS violation probability according to the large derivation theory \cite{wu2003effective}.
We set the value of $\theta$ to be the same for queues in each satellite as an identical QoS requirement.
Due to the deterministic transmission rate, the effective capacity of the queue is defined as \looseness=-1
\begin{equation}
    C(R^{s}_{w, t}, \theta) \triangleq -\frac{1}{\theta}\log\left(\mathbb{E}\left[e^{-\theta R^{s}_{w, t}}\right]\right) = R^{s}_{w, t}.
    \label{eq:eff_capacity}
\end{equation}
By substituting the effective capacity from Eq. \eqref{eq:eff_capacity} into Eq. \eqref{eq:delay_violation} and given the target delay violation probability of the queue $\varepsilon$, we can obtain the queue's delay bound, given by
\begin{equation}
    D^{s, \mathrm{queue}}_{w, t} = -\frac{\log \varepsilon}{\theta R^{s}_{w, t}}.
\end{equation}
Therefore, the overall delay bound of satellite $s$, including the queue's delay bound and the propagation delay, can be represented by
\begin{equation} \label{eq:delay}
    \begin{aligned}
        D^{s}_{w, t} = \left\{
        \begin{aligned}
            & D^{s, \mathrm{queue}}_{w, t} + \frac{d^s_{w, t}}{c}, &&\text{if } R^{s}_{w, t} \not= 0, \\
            &0, &&\text{otherwise},
        \end{aligned}
        \right.
    \end{aligned}
\end{equation}
where $c$ is the light speed.

To satisfy Eq. \eqref{eq:delay_violation}, the effective bandwidth $A(l_{w, t}, \theta)$, which is the minimum required service capability to accommodate the service demand, should be equal to the overall effective capacity from all satellites serving the target area, where
\begin{equation}
    A(l_{w, t}, \theta) \triangleq \frac{1}{\theta}\log\left(\mathbb{E}\left[e^{\theta l_{w, t}}\right]\right).
    \label{eq:effective_bandwidth}
\end{equation}

\subsection{Problem Formulation}
Considering time-varying available resources from satellites covering the target area, we aim to minimize the overall system cost by reserving communication resources from different satellites in each slicing window.
The overall system cost includes resource usage and delay cost.

\subsubsection{Resource Usage}
The resource usage at time $(w, t)$ is the sum of resource consumption from all satellites serving the target area, given by 
\vspace{-1mm}
\begin{equation}
    C^\mathrm{res}_{w, t} = \sum\nolimits_{s \in \mathcal{S}_w} a^s_{w, t} b^{s}_{w} B. \label{eq_res}
    \vspace{-1mm}
\end{equation}

\subsubsection{Delay Cost}
Although there are no specific delay requirements for the delay-tolerant service, users may pursue a lower service delay. 
Let $C^\mathrm{delay}_{w,t}$ denote the delay cost at time $(w, t)$, which notes the maximum overall delay bound from all satellites, given by
\vspace{-1mm}
\begin{equation}
    C^\mathrm{delay}_{w,t} = \max\nolimits_{s \in \mathcal{S}_w} D^{s}_{w, t}. \label{eq_delay}
    \vspace{-1mm}
\end{equation}

To this end, we formulate the resource slicing problem in LSN as 
\vspace{-3mm}
\begin{align}
    \mbox{\textbf{P0} : } \min_{b^{s}_{w}} & \ \sum_{t=1}^{T} \beta_1 C^\mathrm{res}_{w, t} + \beta_2 C^\mathrm{delay}_{w,t}  \label{optimal}\\
    \mbox{s.t.} & \ \sum_{s \in \mathcal{S}_w} C(R^{s}_{w, t}, \theta) \geq A(l_{w, t}, \theta), \tag{\ref{optimal}{a}} \label{op_a}\\
    & \ b^{s}_{w} \in [0, 1], \forall s \in \mathcal{S}_w, \forall w \in \mathcal{W} \tag{\ref{optimal}{b}} \label{op_b}
    \vspace{-2mm}
\end{align}
where $\beta_1$ and $\beta_2$ represent the weight for resource usage and delay cost. Constraint \eqref{op_a} represents that the overall effective capability according to the resource slicing decision for the target area should be larger than the required service capability from the actual service demand.
Constraint \eqref{op_b} limits the resource proportion to be reserved for the slice in each satellite. 

\section{DT-assisted Resource Slicing Scheme in LSN}

\begin{figure}
    \centering
    \includegraphics[width=0.38\textwidth]{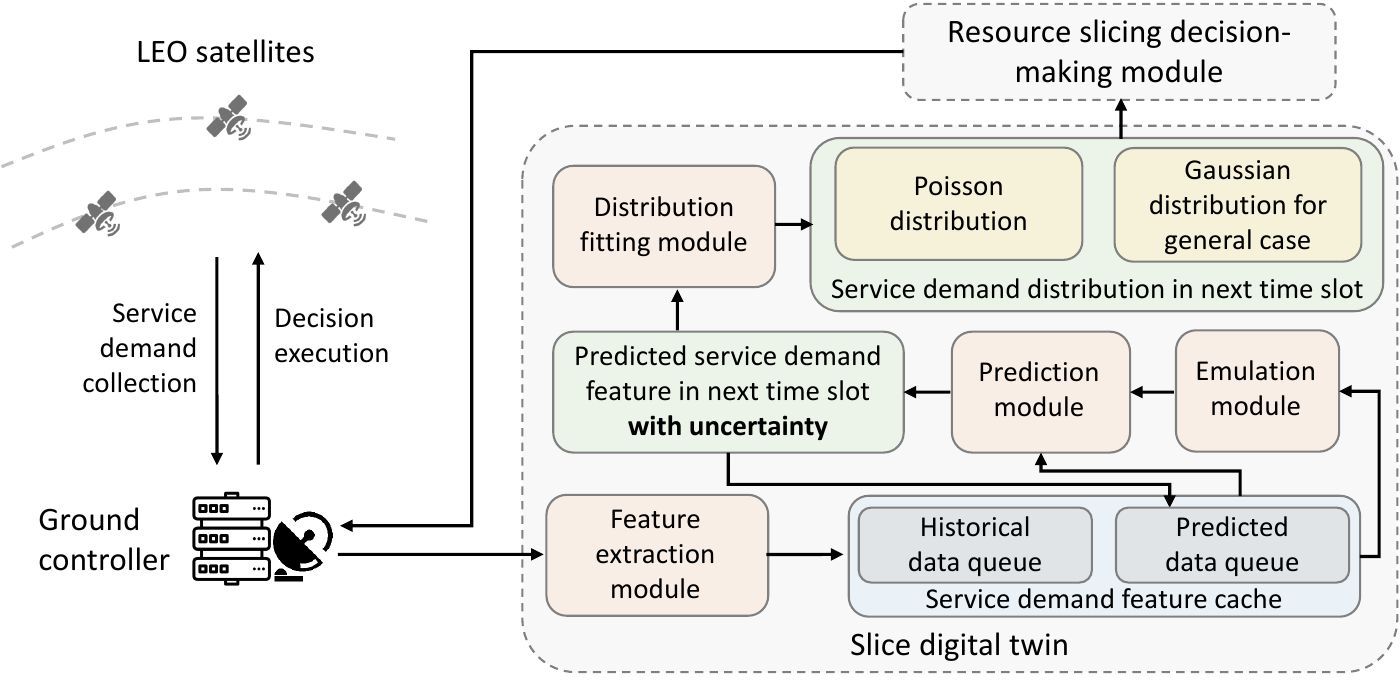}
    \caption{Slice digital twin for LSN.}
    \label{fig:sdt}
\end{figure}

\subsection{Problem Transformation}
The formulated problem $\textbf{P0}$ is a non-linear stochastic optimization problem with temporal dynamic service demands and time-varying available resources due to satellite mobility in LSN.
Solving the problem depends on the predicted service demand for satisfying constraint \eqref{op_a}.
However, obtaining an accurate service demand is challenging due to the non-stationary time-varying service demand and the inaccuracy of the prediction.
Thus, the resource slicing decisions, which heavily rely on accurate service demands, lack robustness and efficiency in a practical environment and result in resource under-/over-provisioning.
Furthermore, the non-stationary and time-varying feature of service demands makes it impractical to directly apply model-free techniques to capture demand variations. \looseness=-1

To tackle the uncertainty of service demands, we transform the original problem by employing chance-constrained programming to provide statistical satisfaction of predicted service demand, thereby achieving a tradeoff between system cost and the risks associated with erroneous predictions, shown as follows: \looseness=-1
\begin{align}
    \mbox{\textbf{P1} : } \min_{b^{s}_{w}} & \ \sum_{t=1}^{T} \beta_1 C^\mathrm{res}_{w, t} + \beta_2 C^\mathrm{delay}_{w,t}  \label{optimal1}\\
    \mbox{s.t.} & \ \mathrm{Pr} \left\{\sum_{s \in \mathcal{S}_w} C(R^{s}_{w, t}, \theta) \geq A(\tilde{l}_{w, t}, \theta) \right\} \geq \gamma, \tag{\ref{optimal1}{a}} \label{op_1a}\\
    & \ \eqref{op_b}, \notag
    \vspace{-2mm}
\end{align}
where $\tilde{l}_{w, t}$ is the predicted service demand following a predicted service demand distribution $\tilde{\mathcal{F}}_{w, t}(\tilde{\mathbf{x}}_{w, t})$ at time $(w, t)$.
Specifically, the parameter $\tilde{\mathbf{x}}_{w, t}$ follows a multivariate Gaussian distribution to represent the prediction uncertainty \cite{atawia2016joint}, and $\gamma \in (0, 1]$ is the satisfaction level of service demand.

\subsection{Slice Digital Twin Establishment}
We propose an adaptive DT-assisted resource slicing (ADTRS) scheme in LSN to enable robust and adaptive resource slicing decisions.
In the ADTRS scheme, we construct a slice DT, which is a digital replica of the network slice of the area to enhance the resource slicing performance, especially in non-stationary environments \cite{qu2024digital, zhou2022digital}.
Specifically, the slice DT is constructed and maintained in the ground controller, which gathers real-time service demands of the area collected from satellites in each time slot as the input data.
In the slice DT, we first design a feature extraction module to extract the service demand feature by processing the collected data.
Then, a prediction module is established to predict service demand features with prediction uncertainty, aiming to capture the non-stationary variation of the service demand.
After that, we devise a distribution fitting module, where two different distributions are considered to fit the service demand distribution and classify the predicted service demand to enhance robust resource slicing.
Moreover, an emulation module is developed, assessing the resource slicing decision for a revised service demand, to estimate the slicing performance, which allows adaptive resource slicing in LSN.
The detailed workflow of the designed slice DT for LSN is presented in Fig. \ref{fig:sdt}, with details of four designed modules explained below:
\subsubsection{Feature extraction module}
To predict the future service demand, the historical service demand is necessary for the predictor.
The ground controller is responsible for collecting the service demand of the target area at the end of each time slot, where the service demand can be sampled by the satellite serving the area, given by $O^s_{w, t} = \{O^s_{w, t, 1}, ..., O^s_{w, t, \tau}\}$.
The sampling frequency is set as 1 per second with $O^s_{w, t, i}$ representing the service demand of the area sampled by satellite $s$ in $i$-th second.
Since sampling is synchronous across all satellites, the sampled service demand for the area can be given by $O_{w, t} = \{O_{w, t, 1}, ..., O_{w, t, \tau}\}$, where $O_{w, t, i} = \sum_s O^s_{w, t, i}$.
Based on the sampled service demand in each time slot, the feature extraction module is designed to capture the service demand feature.
Let $G_{w, t} = U(O_{w, t})$ denote the service demand features of $O_{w, t}$, including the mean and variation of the service demand per second at time $(w, t)$, where $U$ is the feature extraction function mapping the sampled service demand to service demand features.
These features are then added to the historical data queue in the service demand feature cache. \looseness=-1

\subsubsection{Prediction module}
The prediction module predicts the service demand feature at time $(w,t)$, by utilizing the service demand feature in previous $K$ time slots as input.
The averaged service demand feature and the uncertainty can be predicted using the prediction module, given by $(\bar{G}_{w, t}, \rho^G_{w, t}) = g(\xi, G_{w, t-1}, G_{w, t-2}, ...)$, where $\xi$ is the parameter set of the predictor. 
Further details will be provided in the following subsection.
Note that the prediction module is designed for a one-slot prediction to get high accuracy.
However, the controller requires the service demand of the whole slicing window to make the slicing decision at the beginning of the slicing window.
In this case, the averaged predicted feature $\bar{G}_{w, t}$ will be added to the predicted data queue in the service demand feature cache as the input for the prediction of future service demand features. \looseness=-1

\subsubsection{Distribution fitting module}
With the predicted service demand feature and the uncertainty information, we aim to fit the service demand distribution.
This step is crucial for conducting effective capacity analysis and incorporating the stochastic feature of service demand into resource slicing decisions.
Due to the non-stationary service demand variation, it is hard to use a single distribution to model the service demand across different time slots.
In this case, a distribution fitting module is designed to classify the predicted service demand distribution.
In this paper, we adopt two distributions for the fitting: the Poisson distribution, which is commonly encountered in network traffic, and the Gaussian distribution, serving as a general case, as a bunch of diverse traffic tends to conform to this distribution according to the Central Limit Theorem.
By analyzing the predicted feature $\bar{G}_{w, t}$, the service demand at time $(w, t)$ can be classified in one distribution $\tilde{\mathcal{F}}_{w, t}(\tilde{\mathbf{x}}_{w, t})$. \looseness=-1

\subsubsection{Emulation module}
At the beginning of each slicing window, the controller retrieves the predicted service demand feature of each time slot within the slicing window.
Throughout each slicing window, the feature of the actual collected service demand at the end of each time slot is extracted and compared with the predicted feature to obtain the prediction inaccuracy.
Since inaccuracy may accumulate over time slots, the service demand may not be accommodated based on the resource slicing decisions according to the initial prediction.
To address this issue, an emulation module is introduced to emulate the slicing decision for the remaining time slots of the slicing window based on the revised service demand feature according to the actual service demand to achieve an estimated performance. 
Based on the emulation results and leveraging the flexible resource slicing capability in LSN, where slicing decisions are executed only after the satellite covers the area, adjustments can be made to the prediction and slicing decisions to enhance slicing performance. 
Specifically, if emulation results indicate unsatisfied service demand with the original prediction-based slicing decision and there are satellites yet to cover the area, the prediction can be refined and the slicing decision can be adjusted for better performance. \looseness=-1

\subsection{BNN-Based Service Demand Feature Prediction}
To obtain the predicted service demand feature along with its uncertainties, Bayesian Neural Networks (BNN) is a promising approach by extending traditional neural networks in introducing probabilistic distributions over parameters \cite{goan2020bayesian}.
In this work, we combine the long short-term memory (LSTM) networks with BNN to obtain the predicted service demand feature with its uncertainties.
Let $\xi$ denote the parameters of the BNN, which are modeled as distributions to reflect both inherent uncertainty from the service demand variation and the uncertainty from the prediction module.
In this case, given a historical series of service demand features $X$ as the input, a distribution of the output predicted service demand feature, being regarded as following the Gaussian distribution \cite{hernandez2015probabilistic}, can be obtained by sampling BNN parameters, hence capturing the prediction uncertainty. \looseness=-1

To obtain the real distribution of the predicted service demand features, the primary goal in training BNN is to accurately estimate the posterior distribution of the network parameters $P(\xi | X, y)$, where $X$ is the training data and $y$ is the label.
However, it is intractable to directly compute the posterior distribution due to the nonlinearity and high-dimensional parameters.
Variational inference provides a practical solution to approximate the posterior distribution by utilizing a distribution $q(\xi | \nu)$ to approximate the true posterior.
The goal of accurate posterior distribution estimation is then transformed into finding the parameter $\nu$ to minimize the Kullback-Leibler (KL) divergence between $P(\xi | X, y)$ and $q(\xi | \nu)$, represented by
\vspace{-1mm}
\begin{equation}
    \nu^* = \arg\min_\nu \mathrm{KL}[q(\xi | \nu) || P(\xi)] - \mathbb{E}_{q(\xi | \nu)} \log P(y|X, \xi),
    \vspace{-1mm}
\end{equation}
where $P(\xi)$ is the prior distribution of the model.
To this end, leveraging BNN, we can get the predicted service demand feature and uncertainty $(\bar{G}_{w, t}, \rho^G_{w, t})$ through sampling the parameter distribution $P(\xi | X, y)$ to quantify the associated uncertainty of the training data.
In this case, we can finally fit the predicted service demand distribution $\tilde{\mathcal{F}}_{w, t}(\tilde{\mathbf{x}}_{w, t})$ through the predicted feature.
Moreover, periodically retraining through newly collected data can help to capture the non-stationarity of service demand by updating the posterior distribution of the BNN network parameters.

\subsection{DT-Assisted Resource Slicing Algorithm}
With the predicted service demand distribution $\tilde{\mathcal{F}}_{w, t}(\tilde{\mathbf{x}}_{w, t})$, constraint \eqref{op_1a} can be transformed into the following two cases according to the distribution fitting module:
\begin{itemize}
    \item[-] If $\tilde{\mathcal{F}}_{w, t}$ follows Poisson distribution, then the intensity parameter $\tilde{\lambda}_{w, t} \sim \mathcal{N}(\bar{\lambda}_{w, t}, \sigma^2_{w, t})$ is used to describe the distribution, where $\bar{\lambda}_{w, t}$ is the mean value of the intensity and $\sigma^2_{w, t}$ is the variance representing the uncertainty.
    In this case, the constraint can be transformed to
    \begin{align}
        & \mathrm{Pr} \left\{\sum_{s \in \mathcal{S}} C(R^{s}_{w, t}, \theta) \geq \frac{\tilde{\lambda}_{w, t}(e^\theta-1)}{\theta} \right\} \geq \gamma \label{poisson}\\
        \Rightarrow & \  \Phi\left(\frac{\theta \sum_{s \in \mathcal{S}} C(R^{s}_{w, t}, \theta) - \bar{\lambda}_{w, t}(e^\theta-1)}{\sigma_{w, t}(e^\theta-1)}\right) \geq \gamma, \notag
    \end{align}
    where $\Phi(\cdot)$ is the cumulative distribution function of a standard Gaussian distribution.
    \item[-] If $\tilde{\mathcal{F}}_{w, t}$ represents Gaussian distribution, then two random variables $\tilde{\mu}_{w, t}$ and $\tilde{\sigma}^2_{w, t}$ used to represent the mean and variance of the Gaussian distribution. Since the two variables are independent, we have $\tilde{\mu}_{w, t}\sim \mathcal{N}(\bar{\mu}_{\mu, w, t}, \sigma^2_{\mu, w, t})$ and $\tilde{\sigma}^2_{w, t}\sim \mathcal{N}(\bar{\mu}_{\sigma, w, t}, \sigma^2_{\sigma, w, t})$ to depict the average value and the variance of the parameters. 
    Therefore, the constraint can be transformed to
    \vspace{-1mm}
    \begin{align}
        & \mathrm{Pr} \left\{\sum_{s \in \mathcal{S}} C(R^{s}_{w, t}, \theta) \geq \tilde{\mu}_{w, t} + \frac{1}{2} \tilde{\sigma}^2_{w, t} \theta \right\} \geq \gamma \label{gaussian}\\
        \Rightarrow & \  \Phi\left(\frac{\sum_{s \in \mathcal{S}} C(R^{s}_{w, t}, \theta) - (\bar{\mu}_{\mu, w, t} + \frac{1}{2}\theta \bar{\mu}_{\sigma, w, t})}{\sqrt{\sigma^2_{\mu, w, t} + \frac{1}{4}\theta^2 \sigma^2_{\sigma, w, t}}}\right) \geq \gamma. \notag
        \vspace{-1mm}
    \end{align}
\end{itemize}

Then, we aim to find the optimal resource slicing decision for the slicing window.
Considering that the resource slicing decision will be executed only when the satellite first covers the area, let $\tilde{b}^{s}_{w}$ denote the candidate resource slicing decision of problem $\textbf{P1}$.
Since the objective function \eqref{optimal1} includes a conditional term in Eq. \eqref{eq:delay} related to $\tilde{b}^{s}_{w}$, which cannot be directly solved by the optimization technique, we add an additional binary variable $r^{s}_{w}$ to constrain the resource slicing decision and implement the big-M method to deal with the non-convexity under the condition when $\tilde{b}^{s}_{w} = 0$ and $a^{s}_{w, t} = 1$ \cite{lee2011mixed}. 
Therefore, the problem can be transformed into
\begin{align}
    \mbox{\textbf{P2} : } \min_{\tilde{b}^{s}_{w}, r^{s}_{w}} & \ \sum_{t=1}^{T} \beta_1 C^\mathrm{res}_{w, t} + \beta_2 C^\mathrm{delay'}_{w,t}  \label{optimal2}\\
    \mbox{s.t.} & \ \tilde{b}^{s}_{w} \leq r^{s}_{w}, \tag{\ref{optimal2}{a}} \label{op_2a}\\
    & \ r^{s}_{w} \in \{0, 1\}, \tag{\ref{optimal2}{b}} \label{op_2b}\\
    & \ \eqref{op_b}, \eqref{op_1a}, \notag
\end{align}
where $C^\mathrm{delay'}_{w,t} = \max_{s \in \mathcal{S}} D^{\mathrm{M}, s}_{w, t}$ with
\begin{equation}
    D^{\mathrm{M}, s}_{w, t} = -\frac{\log \varepsilon}{\theta \left(C(R^{s}_{w, t}, \theta) + M (1 - r^{s}_{w} a^{s}_{w, t}) \right)} + \frac{d^s_{w, t}}{c}r^{s}_{w} a^{s}_{w, t}, \notag
\end{equation}
and $M$ as a large value.

Since the formulated mixed integer programming problem is a disciplined convex problem, the candidate solution $\tilde{b}^{s}_{w}$ can therefore be directly obtained via convex optimization solvers at the beginning of slicing window $w$.

At the beginning of the slicing window, the executed resource slicing decision $b^{s}_{w}$ is set to 0.
Then, when satellite $s$ is available for the area at time $(w,t)$, given by $\sum_{t'=1}^t a^s_{w, t'} > 0$, the resource slicing decision will be executed by designating $b^{s}_{w} = \tilde{b}^{s}_{w}$.
Due to the flexibility of resource slicing in LSN, the decision can be adjusted in each time slot for satellites whose decisions have not been executed according to the real-collected service demand features.
The slice DT can emulate the current decision to estimate the overall system cost based on the revised predicted service demand feature of the remaining time slots in the slice.
Specifically, an accumulated effect is considered to represent the revised predicted service demand, given by $\bar{G}_{w, t} \gets \bar{G}_{w, t} + \alpha^{t-t'} (G_{w, t'} - \bar{G}_{w, t'})$, where $\alpha$ is a discount factor.
If there still exist satellites not serving the area before time $(w,t)$ in the slicing window, denoted by $\sum_{t'=1}^{t-1} a^s_{w, t'} = 0$ and $\sum_{t'=t}^T a^s_{w, t'} > 0$.
In this case, if the estimated system cost from the emulation is higher than the cost by solving \eqref{optimal2} for the remaining time slots, or constraint \eqref{op_1a} cannot be satisfied, the slice DT will re-predict the service demand feature for the remaining time slots.
Then, the candidate resource slicing decisions for satellites not executed can be updated according to the updated predicted service demand to better adapt to the service demand.
\section{Simulation Results}
In this section, we present simulation results to demonstrate the performance of our proposed ADTRS scheme.
For the LSN, we adopt the Starlink Phase 1 for the satellite constellation parameters, including 72 orbits with 22 satellites in each orbit. 
The inclination angle is 53 degrees and the altitude of the satellite is 550 km. 
By default, we set the elevation angle to be 30 degrees.
The satellite transmit power is set to be 10 dBW, with an antenna gain of 50 dBi and pathloss factor of 2.5.
In the simulation, the target area spans from a latitude of $30^\circ$N to $31.5^\circ$N and a longitude of $82.5^\circ$W to $84^\circ$W, and the service demand is extracted from WIDE Project \cite{cho2000traffic}.
The overall duration of the simulation is 15 min, comprising 90 time slots with each time slot lasting 10 s.
The slicing window length is 10 time slots.
For the LSN service, we consider a content delivery service, where the packet size is set to 10 MB. \looseness=-1

To evaluate the effectiveness of the proposed ADTRS scheme, we adopt the following benchmark schemes: 1) \textit{FRS}: resource slicing decisions are made directly by a deterministic LSTM model without introducing the uncertainty; 2) \textit{FDTRS}: resource slicing decisions are made at the beginning of each slicing window with the assistance of DT, which remains fixed within the slicing window; and 3) \textit{Prefect RS}: the ideal resource slicing strategy that ensures all service demands can be accommodated in the slice with perfect prediction.

\begin{figure}[t]
	\centering
	\subfigure[Overall satellite resource usage]{\includegraphics[width=0.25\textwidth]{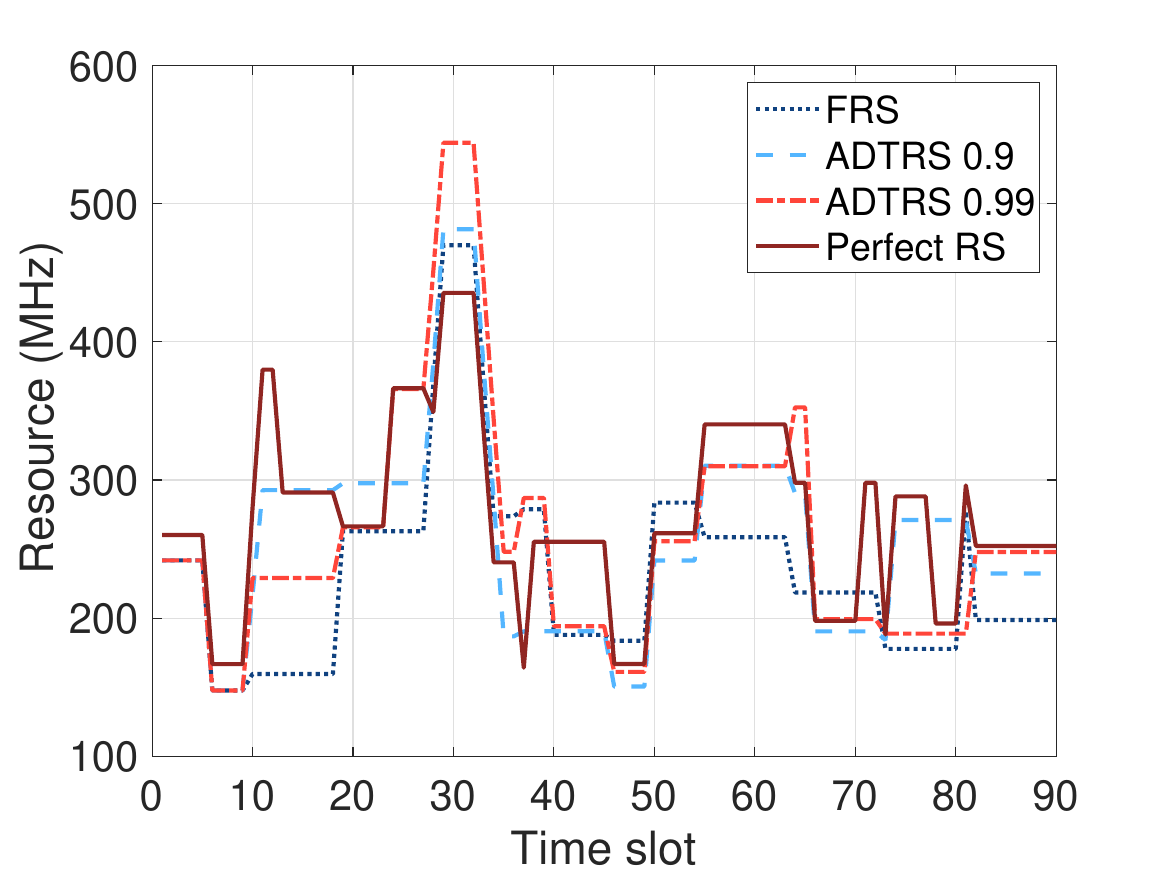}\label{fig:perform_res}}
    \hspace{-5mm}
	\subfigure[Delay cost]{\includegraphics[width=0.25\textwidth]{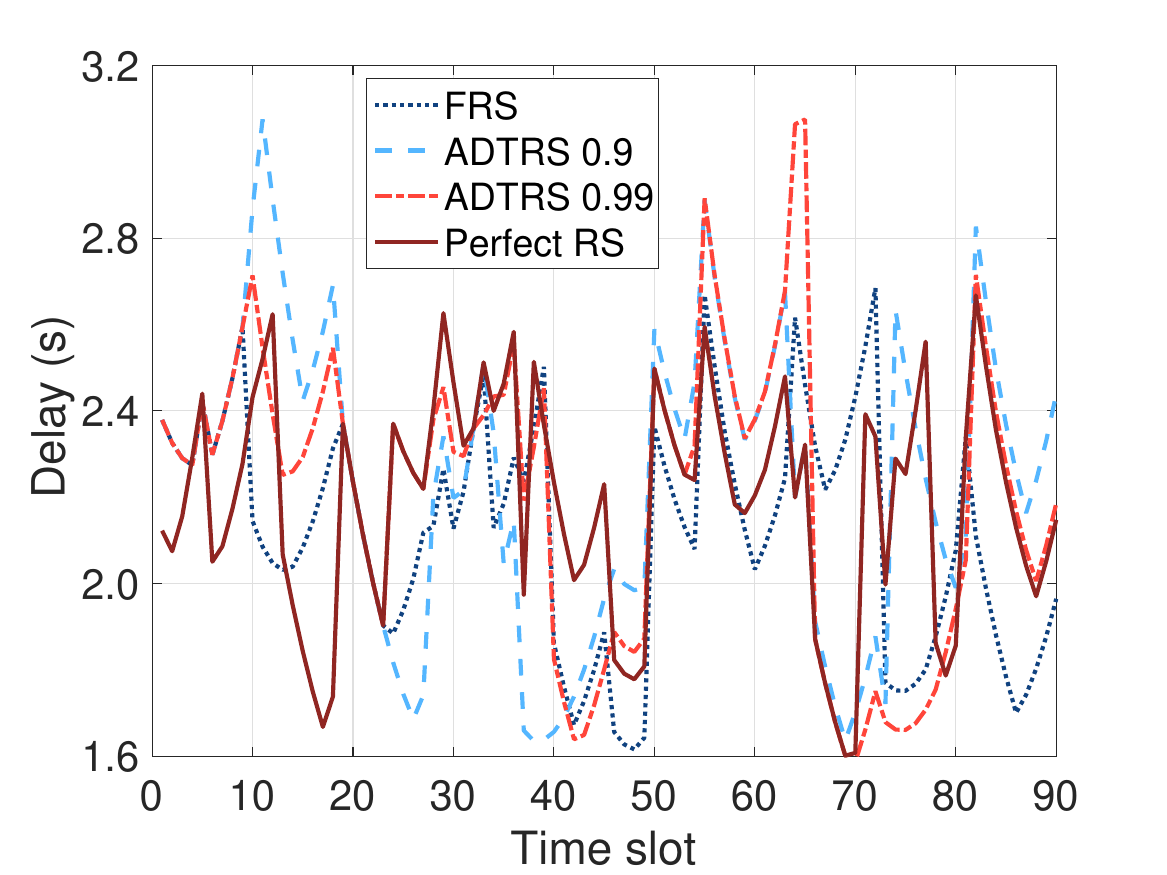}\label{fig:perform_delay}}
	\caption{Performances in each time slot.}
	\label{fig:perform}
\end{figure}

First, in Fig. \ref{fig:perform}, we show the resource usage and delay cost performance over time slots.
From Fig. \ref{fig:perform_res}, we can see that the \textit{FRS} scheme reserves the lowest resource most of the time due to the deterministic service demand prediction.
On the other hand, the ADTRS schemes tend to reserve more resources to adapt to uncertainty from the prediction, resulting in resource usage closer to the \textit{Prefect RS} scheme. 
With a higher satisfaction level (i.e., $\gamma$) of 0.99, more resources are reserved against the large range of uncertainty for supporting more service demands.
Moreover, the ADTRS schemes present an adaptive resource usage compared with \textit{FRS} with more adjustments of decisions, which illustrates the effectiveness of adapting to dynamic service demands through the emulation module in slice DT.
Generally, more reserved resources correspond to lower delay cost, as shown in Fig. \ref{fig:perform_delay}.
To compensate for the under-prediction of service demand, adaptive resource slicing allows for reserving more resources to accommodate service demands; however, reserving additional resources in new satellites will lead to an increase in delay cost, such as the delay cost of ADTRS 0.99 around time slot 63.

\begin{table}[t]
\vspace{1mm}
    \renewcommand\arraystretch{1.02}
    \caption{Statistical performances of different schemes}
    \label{table:perform}
    \centering
    \begin{tabular}{|c|c|c|c|c|}
        \hline
        \multirow{2}*{\textbf{Scheme}} & \textbf{Average} & \textbf{Delay} & \textbf{Violation} & \textbf{Count of}\\
        &\textbf{resource} & \textbf{cost} & \textbf{rate} & \textbf{Re-prediction}\\
        \hline
        \textbf{FRS} & 233 MHz & 2.32 s & 12.2 \% & 0 \\ \hline
        \textbf{FDTRS 0.9} & 242 MHz & 2.20 s & 9.8 \% & 0 \\ \hline
        \textbf{FDTRS 0.99} & 262 MHz & 2.12 s & 6.1 \% & 0 \\ \hline
        \textbf{ADTRS 0.9} & 256 MHz & 2.23 s & 3.7 \% & 40 \\ \hline
        \textbf{ADTRS 0.99} & 260 MHz & 2.20 s & 2.8 \% & 16 \\ \hline
        \textbf{Perfect RS} & 273 MHz & 2.18 s & 0 \% & 0 \\ \hline
    \end{tabular}
\end{table}

The average statistical performances are given in Table \ref{table:perform}.
Specifically, the \textit{FRS} consumes the least resources with the highest average delay, yielding a higher violation rate of service demand.
With the assistance of slice DT, more resources are reserved to reduce the impact of prediction uncertainty for a lower violation rate.
Moreover, adopting adaptive resource slicing, the proposed ADTRS schemes perform the best in violation rate among all schemes by re-predicting the service demand within the slicing window and updating the resource slicing decisions to better fit the actual service demand.
Specifically, a higher satisfaction level, aimed at addressing potentially larger uncertainties, results in fewer re-predictions due to the increased robustness of resource slicing decisions against uncertainty. \looseness=-1

\begin{figure}[t]
\vspace{-5mm}
	\centering
	\subfigure[Average resources in time slot]{\includegraphics[width=0.25\textwidth]{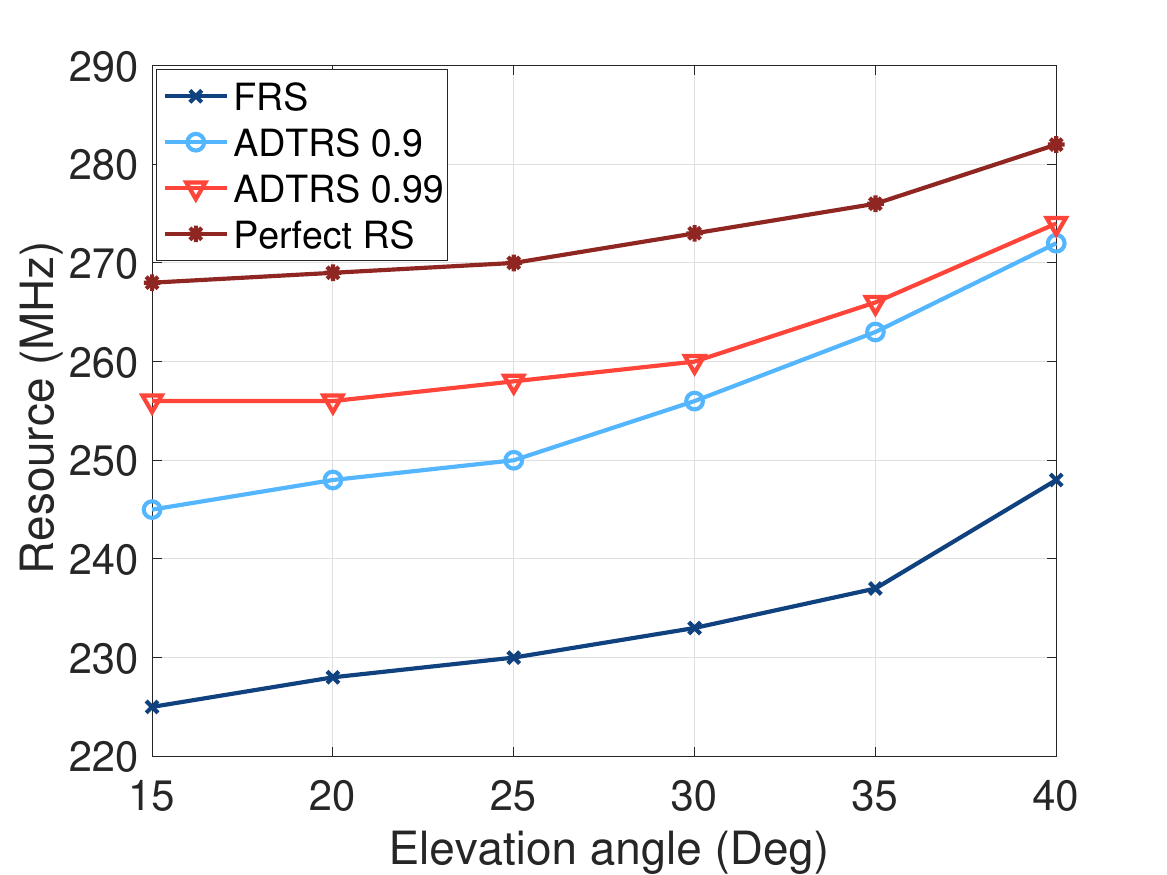}\label{fig:ele_res}}
    \hspace{-5mm}
	\subfigure[Violation rate]{\includegraphics[width=0.25\textwidth]{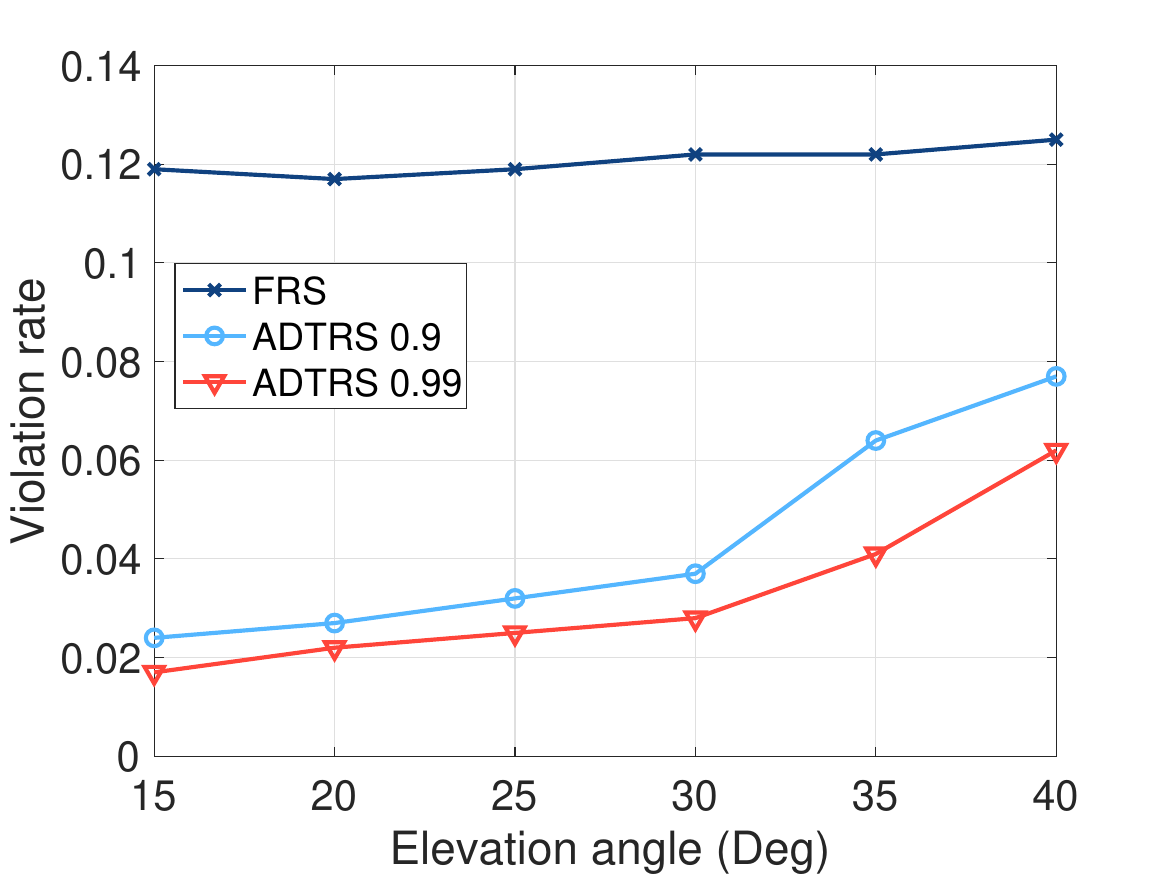}\label{fig:ele_vio}}
 \vspace{-2mm}
	\caption{Performances under different elevation angles.}
 \vspace{-1mm}
	\label{fig:ele}
\end{figure}

Finally, we present the resource consumption and service demand violation rate of our proposed scheme under different minimum satellite elevation angles.
From Fig. \ref{fig:ele}, we can observe that when the elevation angle is small, indicating that more satellites are able to serve the area, less resources from LSN are reserved due to the reasonable combination of resources from different satellites.
The increased number of satellites being available with distinct serving durations allows the ADTRS to showcase its flexibility and adaptiveness in resource slicing, thereby maintaining a low violation rate more effectively.
With the increment of the elevation angle, the resource usage and the violation rate increase accordingly.
The \textit{FRS} scheme has a stable violation rate which is mainly caused by inaccurate prediction, while ADTRS can adaptively reserve resources for robust resource slicing. 

\section{Conclusion}
In this paper, we have proposed an ADTRS scheme leveraging a slice DT to assist resource reservation in LSN.
Specifically, a BNN-based prediction module has been designed in slice DT to capture the prediction uncertainty of the non-stationary service demand, and an emulation module has been developed to allow the adaptive slicing to better satisfy the service demand with low resource consumption.
The proposed scheme can support robust and adaptive resource slicing, capable of accommodating time-varying service demands and satellite mobilities, thus mitigating inefficient resource usage and service demand violations, especially in dynamic network environments.
In the future, we will explore resource slicing among different terrestrial areas in LSN to adaptively manage resources for spatial and service-differentiated slices.

\bibliographystyle{IEEEtran}
\bibliography{Bibliography}

\end{document}